\documentclass{llncs}
\usepackage{amsmath}
\usepackage{float}
\usepackage{threeparttable}
\usepackage{makeidx}  
\usepackage{url}

\begin{document}
\pagestyle{headings}  

\title{Content based Weighted Consensus Summarization}
\titlerunning{Content based Weighted Consensus Summarization}  
%
\author{{Parth Mehta}\inst{1} \and {Prasenjit Majumder}\inst{1}}
\authorrunning{Parth Mehta et al.} 
%
%
\institute{Dhirubhai Ambani Institute of Information and Communication Technology, India\\\{parth\_me,p\_majumder\}@daiict.ac.in}

\maketitle              

\begin{abstract}

Multi-document summarization has received a great deal of attention in the past couple of decades. Several approaches have been proposed, many of which perform equally well and it is becoming increasingly difficult to choose one particular system over another. An ensemble of such systems that is able to leverage the strengths of each individual systems can build a better and more robust summary. Despite this, few attempts have been made in this direction. In this paper, we describe a category of ensemble systems which use consensus between the candidate systems to build a better meta-summary. We highlight two major shortcomings of such systems: the inability to take into account relative performance of individual systems and overlooking content of candidate summaries in favour of the sentence rankings. We propose an alternate method, content-based weighted consensus summarization, which address these concerns. We use pseudo-relevant summaries to estimate the performance of individual candidate systems, and then use this information to generate a better aggregate ranking. Experiments on DUC 2003 and DUC 2004 datasets show that the proposed system outperforms existing consensus-based techniques by a large margin.
\end{abstract}

\section{Introduction}

A plethora of summarization techniques have been proposed in last two decades, but few attempts have been made to combine various summarization techniques to build a meta-summarizer. A study in \cite{hong2014repository} shows that several state-of-art systems with apparently similar performance in terms of ROUGE score, in fact, have very little overlap in terms of content.  Essentially these systems seem to be picking out equally good, but different, information. It is possible to leverage this fact, to build a meta-system that combines all the \emph{good} information across summaries and results in a better coverage. 

A meta-summary can be created either before creating individual summaries or post-summarization. In the first case generally, the ranking algorithm is modified to encompass features from several different summarizers and directly generate the aggregate ranking \cite{mogrenextractive}. In contrast, the latter systems use sentence rankings or summaries generated from individual systems and combine them to form a meta-summary\cite{hongsystem,pei2012supervised,wang2012weighted}. The first type of ensembles depends on carefully combining various aspects of the individual systems, which is not only non-trivial but is also not possible in several cases. In contrast, the second approach can use the existing systems as it is without any modifications, which makes it possible to include as many candidate systems as required, without any overhead. The systems proposed in \cite{mehta2016extractive,mehta2018effective} looks into combining several sentence similarity scores to generate a more robust summary. These approaches show that using various combinations of ranking algorithms and sentence similarity metrics generally outperforms individual systems. In this work, we focus on generating an aggregate ranking of sentences from individual rankings rather than individual summaries. When using just the summaries to generate an ensemble, there is an upper bound on the overall performance\cite{hong2014repository}, since the choice of sentences is limited to the existing summaries rather than entire documents. Both \cite{pei2012supervised} and \cite{wang2012weighted} focus on combining the sentence rankings from candidate systems using weighted linear combinations. While the former relies on a supervised approach that uses SVM-rank to learn relative rankings for all sentence pairs, the latter uses an unsupervised approach based on consensus between the candidate rankings. Existing summarization datasets are too small to train a generic supervised model. In this work, we focus on consensus-based methods to generate aggregates. While our approach is similar in principle to Weighted consensus summarization (WCS)\cite{wang2012weighted}, the way in which we define consensus differs. Unlike WCS, we do not consider sentence rankings to compare two systems. Rather we analyze the overlap in content selected by these systems to measure the consensus between them. We also take into account the relative performance of these systems for individual documents, thus ensuring that best performing system gets more weight compared to the ones with weaker performance.

\section{Consensus based summarization}

Consensus-based summarization is a type of ensemble system that \emph{democratically} selects common content from several candidate systems by taking into account the individual rankings of candidate systems. As opposed to this, the \emph{first past the post} types of ensembles select the highest ranked content from each individual system, even if they are ranked lower in other systems. In case of consensus-based systems, the sentences that are broadly accepted by several systems tend to be ranked higher rather than those championed by only some. Examples are Borda Count\footnote{\url{https://en.wikipedia.org/wiki/Borda_count}}, and Weighted Consensus summarization\cite{wang2012weighted}. Borda count assigns, to each sentence in the original rank lists, a score equal to their rank, i.e. sentence ranked 1$^{st}$ is given a score 1, the one ranked 2$^{nd}$ is given a score 2 and so on. The aggregate score is computed by averaging the score of a sentence in all the rank lists. One major problem with such techniques is their failure to take into account variance in performance of candidate systems across documents. A single system that performs very poorly, can limit the overall performance of the ensemble. 

The weighted consensus summarization\cite{wang2012weighted} creates an aggregate ranking that is as close as possible to the individual rankings. As it is impossible to know beforehand, which candidate system will work best for a given document, the weighted consensus model gives equal importance to all the candidate systems. It then iteratively finds out the aggregate ranking that is as close as possible to each individual ranking. Like other consensus-based methods, WCS fails to take into account the variance in system performance. Another major issue is the manner in which difference between ranked lists is computed. WCS uses L2 Norm to compute the concordance between the aggregate and individual rankings. WCS minimizes equation \ref{eq1} where, r$^*$ is the aggregate rank list, $r_i$ are the individual rankings and $w_i$ are the relative weights assigned to each system. The constraint on $||w||^2$ ensures that the weights are as uniform as possible.
\vspace{-1mm}
\begin{equation}
(1-\lambda)\sum_{i=1}^{K}w_i||\boldsymbol{r^*} - \boldsymbol{r_i}||^2 + \lambda||\boldsymbol{w}||^2
\label{eq1}
\end{equation}

The constraint of minimizing the distance between entire rank lists, instead of the top-k sentences which form the summary, is unnecessary. As long as candidate systems agree in the top-k sentences, which are to be considered for the summary, any additional constraint on lower ranked sentences can adversely affect the performance. Besides that, considering the nature of documents, there will always be more than one sentence which will convey the same information. As a matter of fact, DUC2003 corpus has on average 34 sentences per document cluster, that are repeated at least once, while DUC2004 has 26 such sentences on an average per cluster. There are many more sentences that have near similar information. Simply comparing rank lists of the sentence does injustice, in cases where different sentences with very similar information were selected by different systems. To overcome these two problems, we propose a content-based consensus summarization method, which improves upon the existing WCS method. We use inter-system ROUGE scores to measure the similarity between rankings of two systems. The consensus is then achieved on content, rather than sentence rankings. Under certain constraints, this also takes into account the relative performance of individual systems, when computing the aggregate ranking. The method is described in detail in next section.  
\vspace{-3mm}
\section{Proposed Approach}

As in any consensus-based approach, the idea is to find a weighted combination of individual sentence rankings from the candidate systems to form an aggregate ranking. The problem boils down to finding the best combination of weights that maximizes the ROUGE score.
In the proposed approach we define a new method for assigning weights to different candidate systems. We call this approach \emph{Content based Weighted Consensus Summarization (C-WCS)}. Ideally, a better performing system should contribute more to the aggregate summary compared to a system with lower ROUGE scores. Of course in a practical setup, where the benchmark summaries are not available apriori, it is impossible to know which system will perform better. In theory, it is possible to train a system that can predict this information, by looking at the input document. But in practice, the utility of such a system would be limited by the amount of training data available. Instead of this approach, we propose using \emph{pseudo relevant summaries}. For a given candidate summary $S_i$, each of the remaining $N-1$ candidate systems, {$S_j: j \epsilon\{1...N\}, j\neq i$}, are considered to be \emph{pseudo-relevant} summaries. We then estimate relative performance of the individual system from the amount of content it shares with these \emph{pseudo relevant summaries}. Weights of a candidate system $i$ is computed as shown in equation \ref{eq3}. Sim($S_i$,$S_j$) is defined as ROUGE-1 recall computed considering $S_j$ as the benchmark summary used to evaluate $S_i$.

\begin{equation}
w_i = \frac{1}{N-1}{\sum_{j\neq i}Sim(S_i,S_j)}
\label{eq3}
\vspace{-2mm}
\end{equation}
The underlying assumption in this proposed approach is that the systems performing poorly for a given document are much less in number than the ones performing well. This is not a weak constraint, but we show that this is generally true. In general, a given candidate system tends to perform well on more number of documents compared to the ones on which it performs poorly. Out of the six candidate systems that we experimented with, only one performed below average in more than 30$\%$ cases. The number of documents for which more than 50$\%$  systems performed below average, was 20$\%$. Given this information, we assert that the number of systems performing well for a given document is generally larger than the ones that perform poorly. We present a hypothesis that for a summarization task in general, the \emph{relevant} content in a document cluster is much lower compared to \emph{non-informative} content. Under this assumption, two good or \emph{informative} summaries would have a higher overlap in content, compared to two poor summaries. Simply because the good summaries will have lesser content to choose from, so they are bound to end up with higher overlap. Based on this we argue that the probability of a candidate summary, that has higher overlap with peers, performing better is high.  

The limitation of this approach is the assumption that \emph{good} summaries will have higher overlap amongst themselves, compared to the \emph{bad} summaries. This condition will not be satisfied, if two systems that perform poorly, also generate very similar rankings. But this is not true in general and we show that there is a very good co-relation between rankings generated using Original ROUGE scores (based on handwritten summaries) and the pseudo ROUGE-scores (based on comparison with peers). While the scores themselves differ very much, the system rankings based on these two scores have a Kendal's Tau of 0.7. This indicates that in absence of handwritten summaries, a collection of several peer summaries can serve as a good reference.

\section{Experimental Setup}

The DUC 2003 and DUC 2004 datasets were used for evaluating the experiments. We report ROUGE-1, ROUGE-2 and ROUGE-4 recall. We experiment with six popular and well accepted extractive techniques as the candidate systems for our experiments: Lexrank\cite{erkan2004lexrank}, Textrank, Centroid\cite{radev2004centroid}, FreqSum\cite{nenkova2006compositional}, TopicSum\cite{lin2000automated} and Greedy-KL\cite{haghighi2009exploring}.  We use three baseline aggregation techniques against which the proposed method is compared. Besides Borda Count and WCS, we also compare the results with the \emph{choose-best} Oracle technique. In case of the Oracle method, we assume that the performance of each candidate system, in terms of ROUGE score, is known to us. For each document, we directly select summary generated by the system that scored highest for that particular document and call it the meta-summary. This is a very strong baseline, and average ROUGE-1 score for this meta-system, on the DUC 2003 dataset, was 0.394 compared to a maximum ROUGE-1 of 0.357 for the best performing LexRank system. We further compare the results with two state of the art extractive summarization systems Determinantal Point Processes\cite{kulesza2012determinantal} and Submodular\cite{lin2012learning}. The results are shown in table \ref{tab1} below.
\vspace{-3mm}
\begin{table}[H]
	\begin{threeparttable}
	\centering
	    			\caption{System performance comparison}
			\setlength{\tabcolsep}{1em}
	\begin{tabular}{|c|c|c|c|c|c|c|}
		\hline
		           &                     \multicolumn{3}{c|}{DUC 2003}                     &                \multicolumn{3}{c|}{DUC 2004}                 \\ \cline{2-7}
		  System   &      R-1       &           R-2            &            R-4            &           R-1            &      R-2       &       R-4        \\ \hline
		 LexRank   &     0.357      &          0.081           &           0.009           &          0.354           &     0.075      &      0.009       \\
		 TexRank   &     0.353      &          0.072           &           0.010           &          0.356           &     0.078      &      0.010       \\
		 Centroid  &     0.330      &          0.067           &           0.008           &          0.332           &     0.059      &      0.005       \\
		 FreqSum   &     0.349      &          0.080           &           0.010           &          0.347           &     0.082      &      0.010       \\
		  TsSum    &     0.344      &          0.750           &           0.008           &          0.352           &     0.074      &      0.009       \\
		Greedy-KL  &     0.339      &          0.074           &           0.005           &          0.342           &     0.072      &      0.010       \\ \hline
		  Borda    &     0.351      &          0.080           &          0.0140           &           0.360           &   0.0079             & 0.015 \\
		   WCS     &     0.375      &          0.088           &          0.0150           &          0.382           &     0.093      &      0.0180      \\
		  C-WCS    &     0.390      & \textbf{0.109}$^\dagger$ &          0.0198           & \textbf{0.409}$^\dagger$ & \textbf{0.110} & \textbf{ 0.0212} \\
		  Oracle   & \textbf{0.394} &          0.104           & \textbf{0.0205}$^\dagger$ &          0.397           &     0.107      &      0.0211      \\
		Submodular &     0.392      &          0.102           &          0.0186           &          0.400           & \textbf{0.110} &      0.0198      \\
		   DPP     &     0.388      &          0.104           &          0.0154           &          0.394           &     0.105      &      0.0202      \\ \hline
	\end{tabular}
    \label{tab1}

	\begin{tablenotes}
	\small
	\item Figures in bold indicate the best performing system
	\item ${^\dagger}$ indicates significant difference with $\alpha=0.05$
	\end{tablenotes}
\end{threeparttable}
\end{table}
\vspace{-3mm}
In all cases, the proposed C-WCS system outperforms other consensus-based techniques, Borda and WCS by a significant margin. It performs at par with the current state of art Submodular and DPP systems. In several cases, C-WCS even outperformed the Oracle system, which relies on apriori knowledge about which system will perform the best. We conducted a two-sided sign test to compare the C-WCS system with other systems. $^\dagger$ indicates that the best performing system is significantly better than the next best performing system.  

\section{Conclusion}

In this work, we propose a novel method for consensus-based summarization, that takes into account content of the existing summaries, rather than the sentence rankings. For a given candidate summary we treat other peer summaries as pseudo relevant model summaries and use them to estimate the performance of that candidate. Each candidate is weighted based on their expected performance when generating the meta-ranking. The proposed C-WCS system outperforms other consensus-based aggregation methods by a large margin and performs at par with the state-of-art techniques. 
\bibliographystyle{splncs}
\bibliography{llncs.bib}

\begin{thebibliography}{10}

\bibitem{hong2014repository}
Hong, K., Conroy, J.M., Favre, B., Kulesza, A., Lin, H., Nenkova, A.:
\newblock A repository of state of the art and competitive baseline summaries
  for generic news summarization.
\newblock In: Proceedings of Language Resources and Evaluation Conference.
  (2014)  1608--1616

\bibitem{mogrenextractive}
Mogren, O., K{\aa}geb{\"a}ck, M., Dubhashi, D.:
\newblock Extractive summarization by aggregating multiple similarities.
\newblock In: Proceedings of Recent Advances In Natural Language Processing.
  (2015)  451--457

\bibitem{hongsystem}
Hong, K., Marcus, M., Nenkova, A.:
\newblock System combination for multi-document summarization.
\newblock In: Proceedings of the 2015 Conference on Empirical Methods in
  Natural Language Processing, Lisbon, Portugal, Association for Computational
  Linguistics (September 2015)  107--117

\bibitem{pei2012supervised}
Pei, Y., Yin, W., Fan, Q., Huang, L.:
\newblock A supervised aggregation framework for multi-document summarization.
\newblock In: Proceedings of 24th International Conference on Computational
  Linguistics: Technical Papers. (2012)  2225--2242

\bibitem{wang2012weighted}
Wang, D., Li, T.:
\newblock Weighted consensus multi-document summarization.
\newblock Information Processing \& Management \textbf{48}(3) (2012)  513--523

\bibitem{mehta2016extractive}
Mehta, P.:
\newblock From extractive to abstractive summarization: {A} journey.
\newblock In: Proceedings of the {ACL} 2016 Student Research Workshop, Germany,
  ACL (2016)  100--106

\bibitem{mehta2018effective}
Mehta, P., Majumder, P.:
\newblock Effective aggregation of various summarization techniques.
\newblock Information Processing \& Management \textbf{54}(2) (2018)  145--158

\bibitem{erkan2004lexrank}
Erkan, G., Radev, D.R.:
\newblock Lexrank: Graph-based lexical centrality as salience in text
  summarization.
\newblock Journal of Artificial Intelligence Research \textbf{22} (2004)
  457--479

\bibitem{radev2004centroid}
Radev, D.R., Jing, H., Sty{\'s}, M., Tam, D.:
\newblock Centroid-based summarization of multiple documents.
\newblock Information Processing \& Management \textbf{40}(6) (2004)  919--938

\bibitem{nenkova2006compositional}
Nenkova, A., Vanderwende, L., McKeown, K.:
\newblock A compositional context sensitive multi-document summarizer:
  exploring the factors that influence summarization.
\newblock In: Proceedings of the 29th annual international ACM SIGIR conference
  on Research and development in information retrieval, ACM (2006)  573--580

\bibitem{lin2000automated}
Lin, C.Y., Hovy, E.:
\newblock The automated acquisition of topic signatures for text summarization.
\newblock In: Proceedings of the 18th conference on Computational
  linguistics-Volume 1, Association for Computational Linguistics (2000)
  495--501

\bibitem{haghighi2009exploring}
Haghighi, A., Vanderwende, L.:
\newblock Exploring content models for multi-document summarization.
\newblock In: Proceedings of Human Language Technologies: The 2009 Annual
  Conference of the North American Chapter of the Association for Computational
  Linguistics, Association for Computational Linguistics (2009)  362--370

\bibitem{kulesza2012determinantal}
Kulesza, A., Taskar, B.,  et~al.:
\newblock Determinantal point processes for machine learning.
\newblock Foundations and Trends in Machine Learning \textbf{5}(2--3) (2012)
  123--286

\bibitem{lin2012learning}
Lin, H., Bilmes, J.:
\newblock Learning mixtures of submodular shells with application to document
  summarization.
\newblock In: Proceedings of the Twenty-Eighth Conference on Uncertainty in
  Artificial Intelligence, AUAI Press (2012)  479--490

\end{thebibliography}
\end{document}